\documentclass{article}
\usepackage{arxiv}

\usepackage{booktabs}
\usepackage{array}
\usepackage[english]{babel}
\usepackage{amssymb}
\usepackage[fleqn]{amsmath}
\usepackage[table]{xcolor}
\usepackage{color, colortbl}
\usepackage[pdftex]{graphicx}
\usepackage{multirow}
\usepackage{enumerate}
\usepackage{adjustbox}
\usepackage[mathlines]{lineno}
\usepackage[hidelinks]{hyperref}
\usepackage{tikz}
\usepackage{tabularx}
\usepackage{afterpage}
\usepackage{csquotes}
\usepackage[ruled,vlined,linesnumbered]{algorithm2e}
\usepackage[numbers,sort&compress]{natbib}
\usepackage{nameref}
\usepackage{url}
\usepackage{float}

\MakeOuterQuote{"}

\graphicspath{{../Figures/}{Figures/}}

\title{Hybrid Work and the Restructuring of Urban Mobility in U.S. Cities}

\author{
Baoqi Eileen Chen\textsuperscript{1} \and
Marta C. Gonz\'{a}lez\textsuperscript{1,2,3}
}

\date{
\textsuperscript{1}Department of City and Regional Planning, University of California, Berkeley, CA, USA\\
\textsuperscript{2}Department of Civil and Environmental Engineering, University of California, Berkeley, CA, USA\\
\textsuperscript{3}Energy Technologies Area, Lawrence Berkeley National Laboratory, Berkeley, CA, USA
}

\begin{document}

\maketitle

\begin{abstract}
Entering the post-pandemic era, cities navigate a new normal shaped by hybrid work and space-time flexibility,
but existing evidence is spatially scattered and temporally limited.
Here we develop an analytical approach that examines this reorganization through three dimensions:
remote work adoption via sector composition, daily travel behavior via trip-level metrics, and functional urban form via mobility-derived measures.
To capture the interplay of urban dynamics, we synthesize three complementary spatial metrics.
In particular, coupling overall trip centrality with work-trip concentration differentiates whether employment and daily activity organize around the same centers.
Applied to population-scale data across 15 U.S. metropolitan areas spanning 2019–2024,
our approach shows that daily car travel increased despite elevated remote work,
a pattern consistent across cities with distinct spatial structures.
Decomposition and regression identify trip frequency as the primary contributor to VKT growth, moderated by compact urban form and concentrated employment.
The approach provides a replicable framework for comparative study of post-pandemic urban mobility.
\end{abstract}

\section{Introduction}
\label{sec:intro}

Human behaviors are remarkably consistent, shaped by routines, habits, and social norms that change only gradually over time.
The COVID-19 pandemic disrupted this inertia, constituting a global behavioral experiment that restructures the ways we live, work, and move around \cite{hunter2024city}.
Under nationwide lockdowns in 2020, economic activities halted \cite{bartik2020impact,guan2020global}, social interactions declined \cite{liu2021rapid}, and emissions plummeted \cite{lequere2020co2,venter2020pollution}.
While aggregate indicators rebounded during phased reopenings throughout 2021\cite{stock2025recovering, nyt2025charts}, notable behavior shifts persist.
Stickiness of work-from-home (WFH)  \cite{aksoy2025global} challenged downtown recovery \cite{forouhar2024dtrecovery,leong2023dtrecovery}, accelerating a ``donut effect'' of decentralized urban activity \cite{bondsmith2025shadowdonut,ramani2024wfhdonut}.
This flexibility redistributes activities across space and time.
Commuting is structurally weakened as a daily anchor, being partially replaced by home-based work \cite{restrepo2022wfhatus}, e-commerce \cite{szasz2022ecommerce}, and expanded use of ``third places'' \cite{li2024thirdplace,caros2022flexworkloc}.
People spent more time outside home at non-work locations on WFH days \cite{bouzaghrane2024tele}.
Leisure time rose among workers in remote-friendly jobs \cite{makridis2025atus}, along with a drop in Friday work hours \cite{makridis2025sync}.
As societies transition into the post-pandemic era, a ``new normal'' shaped by hybrid work and flexibility is beginning to take shape.

Empirically characterizing new normal behaviors matters as they underlie transport demand forecasts, downtown recovery strategies, and emissions trajectories.
Establishing a descriptive baseline is a prerequisite for evaluation and intervention.
However, this post-pandemic new normal remains poorly characterized.
Existing knowledge is spatially narrow, often confined to case-study cities, and temporally limited, with most studies focused on disruption and reopening.
In particular, we lack long-term evidence on how mobility evolves under persistent restructuring beyond temporary shocks.
Changes in when and where people work, visit, and socialize manifest in movements through space.
Understanding daily mobility patterns, and how they generalize or vary across urban contexts, therefore provides a direct lens onto post-pandemic urban dynamics.

To this end, naturally-occurring data streams from mobile devices have enabled observations of human mobility behavior in precision and scale \cite{du2025humobreview,batty2010pulse}, revealing regularities in daily movement \cite{gonzalez2008understanding,schneider2013unravelling,ccolak2016understanding} and predictability of individual trajectories \cite{song2010limits}.
These characteristics allow models leveraging mechanistic \cite{jiang2016timegeo, song2010model}, activity-based \cite{tozluoglu2025model, bassolas2019matsim}, and generative approaches \cite{cao2026dtg, yuan2026worldmove, kashiyama2023pflow,pozdnoukhov2018model} to reproduce travel demand from sparse passive inputs alone, and with greater fidelity when fused with survey, land-use, or ground-truth data.
During the pandemic, passive data enabled population-level monitoring of disruptions \cite{lucchini2021livepandemic,oliver2020mobcovid}, and mobility network models derived from such data served as analytical inputs for simulating reopening strategies and predicting infection disparities \cite{chang2021mobreopen}.
As cities experienced and emerged from COVID-19, location-based service (LBS) data captured the restructuring of daily mobility.
After lockdown restrictions lifted, people's spatial mobility recovered faster than temporal coordination  \cite{santana2023covidcoord}.
Spatial exploration rates returned to pre-pandemic levels in 2021 along with reinforced schedule similarities \cite{bouzaghrane2024humobreshp}.
Workers made more weekly trips but visited fewer unique locations \cite{bouzaghrane2024tele},
which corroborates a lasting decrease in diversity of urban physical encounters between 2020 and 2021 due to pandemic-lowered willingness for social exploration \cite{yabe2023diversity}.
Though levels of segregation and isolation returned to 2019 for many cities in 2022 \cite{renninger2025ringpocket}, downtowns had less mixing than before.

Among mobility outcomes, vehicle kilometers traveled (VKT) offers an integrative measure of individual mobility under the new normal \cite{concas2024covidvmt,carb2021postcovidvmt,xiao2023vibrancy,kane2025covidvmtlu}.
VKT measures the total distance driven by all vehicles in a region, aggregating trip frequency, travel distance, and mode choice into a single behavior outcome.
VKT is socially consequential, long used to assess congestion\cite{duranton2011law}, infrastructure \cite{ecola2012fhwa}, and emission \cite{axsen2020ghgpolicy}.
Reducing vehicular carbon emissions is essential for climate change mitigation in urban mobility.
In the U.S., motor vehicles emit over $80\%$ of transportation $CO_2$ \cite{cbo2022trptemsn}, and urban areas contribute about $80\%$ of vehicle $CO_2$ emissions growth \cite{gately2015citytrafco2}.
Accordingly, VKT has been examined as an outcome of ICT-enabled activity substitution and telecommuting \cite{mokhtarian2006ict, mokhtarian1998tele, salomon1986tele}, with renewed attention following the COVID-19 shift to remote work \cite{shen2025remote,wang2024televmt}.

We develop an analytical framework that examines the post-pandemic reorganization of work and daily mobility through three dimensions:
remote work adoption via sector composition, daily travel behavior via trip-level metrics, and functional urban form via mobility-derived measures.
To capture the interplay of urban dynamics, we synthesize three complementary spatial metrics.
In particular, comparing overall trip centrality with work-trip concentration shows whether employment and daily activity organize around the same centers.
We apply the approach to a multi-year, LBS-calibrated mobility dataset comprising billions of trip records across 15 U.S. combined statistical areas (CSA) (see Supplementary Fig.~S5, Tables S4 and S5), spanning eight seasons between 2019 and 2024.
To our knowledge, this constitutes a novel longitudinal, comparative analysis of its kind.
Data are provided by Replica and generated via a nationwide activity-based travel demand model \cite{ReplicaSeasonalMobility,pozdnoukhov2018model} (see ``\nameref{sec:methods}'').
For demographic realism, a comprehensive synthetic population is constructed from Census PUMS, BLS, CTPP, and LEHD;
for behavioral realism, travel behavior personas are learned from multiple proprietary sources including mobile phone location signals and in-dash vehicle GPS, covering approximately 30 million devices per month.
Remote work status is integrated by combining administrative surveys with passive data-inferred home-work locations and activity patterns.
Daily trajectories are simulated on real transport networks.

Across the 15 urban areas studied, our approach shows that remote work remains well above pre-pandemic levels,
though sector composition alone does not account for the variation.
Nonetheless, daily car travel increased despite reduced commuting.
Trip-level metrics show increased frequency as the primary contributor,
accompanied by contracted activity spaces and incomplete transit recovery.
Surprisingly, these patterns hold across cities that range widely in compactness, mobility centrality, and employment concentration.
Differentiating work attractors within monocentric and polycentric spatial structures further disentangles commute and non-commute sources of travel demand.
This urban form characterization makes the reorganization of work and travel interpretable and comparable across multiple cities.
Regression shows mobility behavior and urban form are jointly associated with VKT change, with compactness and employment concentration partially mitigating the increase.
Our proposed approach is replicable for comparative post-pandemic urban mobility analysis.

\section{Results}

\subsection{Remote work adoption reflects local economic structure}
Remote work remains elevated but varied across metropolitan areas, following similar temporal trends but diverging in the specific levels.
Fig. \ref{fig: wfh}A shows that WFH rates surged in Spring 2021, then gradually declined and stabilized around $20\%$, which is about three times higher than 2019 ($\sim 5\%$).
By 2024, Phoenix, Atlanta, and San Jose had the highest WFH rates ($\sim 25\%$), compared with Cleveland and Houston ($\sim 15\%$).
This signals that local economic factors impact remote work adoption beyond immediate pandemic responses.
Fig. \ref{fig: wfh}B traces employment in CSAs through industry sectors to the mode of work.
Large CSAs share similar employment structures, dominated by healthcare, retail, professional services, and manufacturing (see Supplementary Fig. S7 for top sector breakdown in each CSA).
Yet, Fig. \ref{fig: wfh}C exposes a nuance: sectors with the highest remote work adoptions ($35-40\%$), such as Information, Finance, Science/Tech, constitute relatively small employment shares, typically just $2-9\%$. Most workers occupy moderate-to-low WFH sectors.

To further understand variations in WFH adoption, we compare the CSAs by their job diversity and dominant sector.
We measure job diversity with Shannon entropy of employment percentages across the North American Industry Classification System (NAICS) sectors, with higher entropy indicating more balanced employment in varied industries.
As shown in Fig. \ref{fig: wfh}D, diversified economies appear to facilitate workplace flexibility.
WFH rates are lower in Cleveland and Detroit, where employments are concentrated in fewer sectors.
However, this relationship shows mixed signals.
The dominant sector adds additional contexts, although variation is limited by broad sectoral similarity.
San Jose's concentration in Science/Tech corresponds with high WFH adoption ($26\%$), while Detroit's Manufacturing dominance aligns with lower rates ($20\%$).

\subsection{More frequent, shorter car trips underlie daily VKT growth}
Persistent remote work adoption reduced commuting.
Still, VKT increased in nearly all CSAs.
Fig. \ref{fig: raincloud}A traces this increase:
daily VKT per capita had exceeded pre-pandemic baselines by Fall 2024, with most urban areas experiencing around 5-km increase.
Temporal patterns show that VKT displayed the greatest variations in 2021 as cities reopened after lockdowns.
However, from 2022 onward, VKT gradually stabilized and consistently increased.
This observation holds across CSAs with different VKT baseline values (ranging 20-40 km), suggesting systematic behavior changes.
We investigated this observation with behavior outcome metrics central to transportation planning \cite{cervero2001betrav} and human mobility \cite{gonzalez2008understanding} literatures.

Decreased transit use partially accounts for higher VKTs.
Fig. \ref{fig: raincloud}B shows that daily transit distance (km) traveled collapsed in 2021 and did not fully recover, remaining 0-1 km per capita lower than pre-pandemic through 2024.
The effect is pronounced in transit-oriented CSAs (New York, Chicago, San Jose, Philadelphia, and Boston), with New York experiencing about $60-70\%$ drop in 2021.
The mode shift from transit toward cars contributed to more vehicle travel, but the magnitude of transit loss alone cannot fully account for the VKT growth observed.

The largest contributor is increased car trip frequency, as shown in Fig. \ref{fig: raincloud}C.
On average, individuals make about one additional car trip every two days.
The number of daily car trips displayed greatest variance in 2021, suggesting effects of reopening.
Then it continuously increased throughout 2024, indicating widespread change in how people structure daily activities around automobile use.
The increase in trip frequency is accompanied by observations of contracted daily activity space and shorter average trip distances (See Supplementary Fig. S8).
Fig. \ref{fig: raincloud}D shows that daily activity radii consistently contracted by about 0-1 km below baseline throughout 2023, after an initial recovery in Spring 2021. Signs of expansion appeared in 2024.

\subsection{Ubiquitous mobility behavior across diverse urban forms}
This restructuring of mobility appears across all 15 CSAs spanning a diverse urban form spectrum.
To unpack this ubiquity, we characterize each city's spatial structure through three complementary metrics that differentiate how urban form relates to travel patterns.
In doing so, we present a behavior-centric computational framework for measuring urban form via mobility data.

The Population Gini Index ($PopGini$) captures compactness: higher values indicate concentrated settlement, which typically reduces travel, while lower values signal sprawl.
The $\Delta$KS statistic, proposed by Xu et al. \cite{xu2023urbdyn}, measures mobility centrality through the differentiation of daily activity radius ($R_g$) distribution across residents living in 3-km rings from downtown (the 0-3 km ring). High $\Delta$KS values indicate trips converging on a central business district (CBD), while low values imply similar $R_g$ distributions and dispersed destinations.
We also adopt the benchmarks for compactness and monocentricity from Xu et al. \cite{xu2023urbdyn} based on 21 cities around the world, under which U.S. cities are primarily dispersed-polycentric (See Methods and Supplementary Fig. S9).
To reveal how much a city's daily travel organizes around work, we integrate the Urban Centrality Index ($UCI$) proposed by Pereira et al. \cite{pereira2013uci} alongside the $\Delta$KS statistic.
UCI extends the spatial separation index \cite{mk2002vi} with a centrality scale comparable across different geographic areas.
Studies have calculated $UCI$ using static data including population \cite{fan2022equality,xu2020law}, points of interests \cite{rajput2024flood, ma2024flood, fan2023ml}, and jobs \cite{garcialopez2018income}.
To capture the trip attractor effect of work locations, we implement counts of home-to-work trips per census block group (CBG) to quantify employment concentration through individual travel.

\begin{figure}[htbp]
    \centering
    \includegraphics[width=\textwidth]{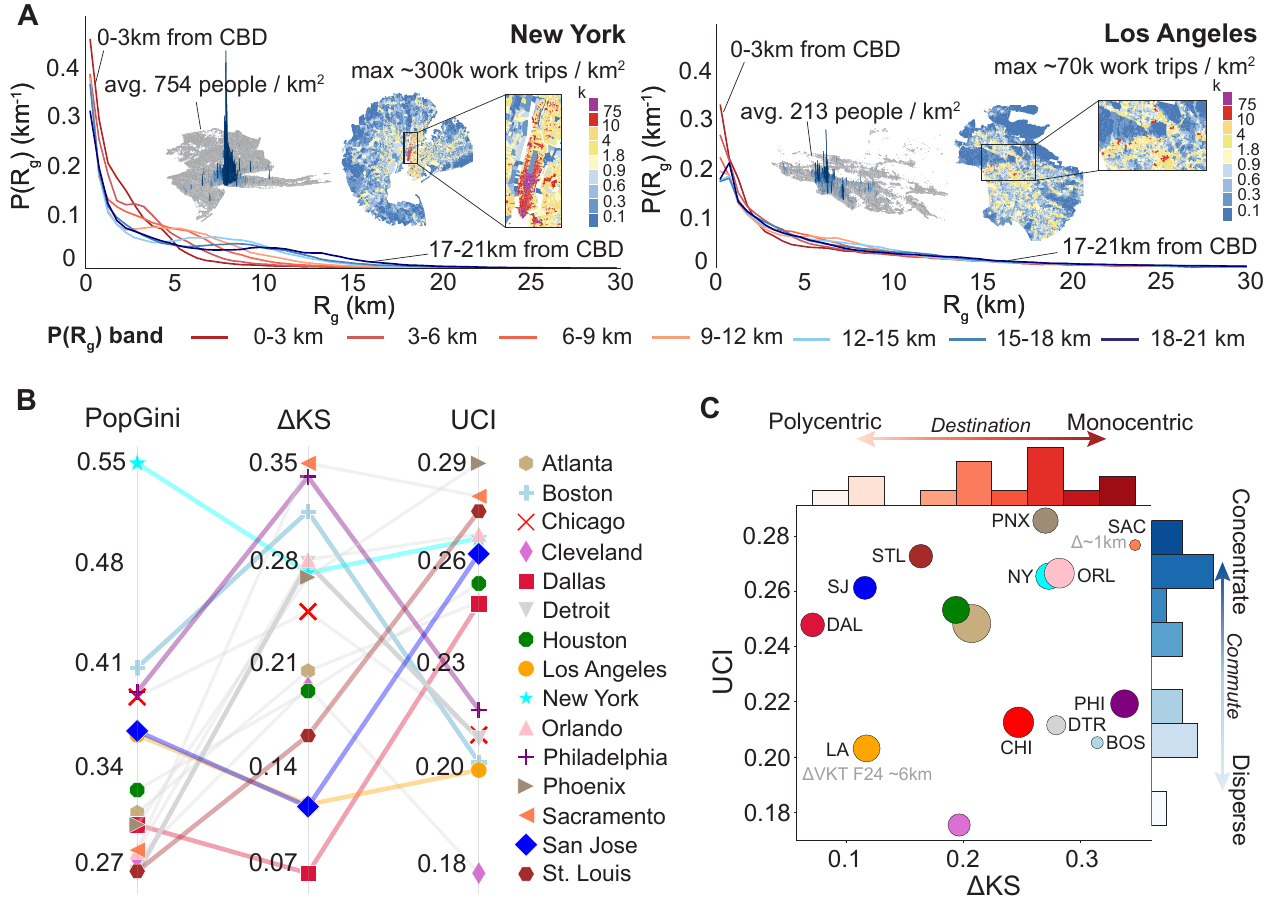}
    \caption{{\bf Interplay of three urban form and function factors in 15 CSAs.} $PopGini$ is computed from 1-km grid-level population counts in 2019, reflecting density as compact (high $PopGini$) or dispersed (low). $\Delta$KS and $UCI$ are computed from baseline mobility data in Fall 2019, using all trips and work trips, respectively, to capture mobility centrality (high $\Delta$KS, monocentric) and employment concentration (high $UCI$, clustered job centers). \textbf{A.} features New York and Los Angles as two characteristic cases of ``compact, monocentric, work attractor'' and ``dispersed, polycentric, distributed work locations''. Distribution plots show probability density functions of residents' $R_g$ by 3-km rings from downtown ($r_0$) based on home location. Darker the reds, closer to downtown (0-3 km); darker the blue, further in suburbs (17-21 km). Population density plots visualize average people in each 1-km grid. Work trip density maps show census-block-group (CBG) level home-to-work trip density. \textbf{B.} plots three metrics for all 15 CSAs in parallel, descending by magnitude panels, highlighting cases where their interplay reveals interesting spatial structures. \textbf{C.} plots $\Delta$KS and $UCI$ in 2D, scatters sized by Fall 2024 VKT change compared with Fall 2019 baseline, differentiating the effect of destination pull and commute anchoring.}
    \label{fig: form}
\end{figure}

Fig. \ref{fig: form} displays the urban form spectrum spanned by the metropolitan areas.
New York highlights one archetype:
Residents live close together in compact settlements ($PopGini$ = 0.546) and relatively monocentric structure ($\Delta$KS = 0.273) compared with other CSAs, along with notable employment concentration ($UCI$ = 0.265) where Manhattan serves as the main work trip attractor.
Fig. \ref{fig: form}A visualizes it spatially.
Population density spikes near the urban core.
Long-tail $R_g$ distributions suggest that outer-ring residents have larger daily activity radii than those living near the CBD, reflecting movement to concentrated downtown destinations.
This aligns with work trip distributions, with steep density gradients from Manhattan outward, characteristic of monocentric cities where peripheral residents travel greater distances to access job opportunities.
Los Angeles characterizes the opposite type.
Low population compactness ($PopGini$ = 0.360) compared with New York ($PopGini$ = 0.546) suggests urban sprawl with car dependence.
Low mobility centrality ($\Delta$KS = 0.118) reflects polycentric destinations, along with low employment concentration ($UCI$ = 0.203) implying scattered work trips to multiple employment centers.

Connecting the three metrics portrays the complex interplay of urban spatial structure (Fig. \ref{fig: form}B).
For example, San Jose has population dispersion ($PopGini$ = 0.363) and polycentric mobility ($\Delta$KS = 0.116) that are similar to Los Angeles (0.360, 0.118),
but its higher employment concentration ($UCI$ = 0.261, close to New York), shaped by Silicon Valley's tech campuses, creates different travel demands.
On the other hand, while Boston and Philadelphia resemble New York's compact-monocentric profile with high $PopGini$ (0.406, 0.390) and $\Delta$KS (0.314, 0.338) values,
their work trip destinations are more spread out, with $UCI$ values (0.205, 0.219) close to Detroit (0.211) and Los Angeles (0.203).
Dallas and St. Louis display an interesting decoupling:
being overall polycentric ($\Delta$KS = 0.072, 0.164) despite moderate-to-high (relative to the CSAs studied) employment concentration ($UCI$ = 0.248, 0.273).

In particular, the coupling of $\Delta$KS and $UCI$ reveals how destination pull (the tendency for trips to converge on downtown cores) interacts with commute anchoring (the degree to which employment locations concentrate or disperse work trips).
We observe a bimodal distribution with four archetypes (Fig. \ref{fig: form}C).
Los Angeles characterizes polycentric--dispersed work, opposite to monocentric--concentrated work,
where Orlando, New York, Phoenix, and Sacramento reflect strong downtown pull for both commute and non-commute trips.
Unlike Los Angeles, other polycentric CSAs display workplace concentration:
in Dallas, San Jose, and St. Louis, employment clusters pull commutes inward even as overall trip-making remains decentralized.
For Boston, Chicago, and Detroit, while all-purpose travel tends to centralize,
employment sites are sufficiently spread to weaken commute concentration, shaping monocentric--dispersed work patterns.
These distinctions differentiate how post-pandemic VKT shifts manifest across different urban forms.

\subsection{Role of mobility behavior and urban form in VKT growth}
The preceding analyses characterize mobility shifts and urban form independently.
To quantify their relative contributions,
we model VKT change (vs. Fall 2019 baseline; $N$ = 105) as a function of mobility behavior shifts, baseline urban form, and baseline VKT.
The resulting model explains about $32\%$ of VKT growth variance ($R^2 = 0.32$),
with daily activity radius ($\Delta R_g$), trip frequency ($\Delta$ Trip), compact urban form ($PopGini_{F19}$), and employment concentration ($UCI_{F19}$) showing significant associations.
Grouped Lindeman–Merenda–Gold (LMG) decomposition \cite{Lideman1980} attributes $51\%$ of explained variance to mobility behavior, $28\%$ to baseline urban form, and $21\%$ to baseline VKT (Fig. \ref{fig: regression}B).
Detailed statistics are provided in Supplementary Tables S6-9.

Fig. \ref{fig: regression}A displays raw regression coefficients along with each variable's relative importance.
Within mobility behavior, increased trips ($\Delta$Trip) are significantly associated with VKT growth,
with each additional trip per day per capita associated with roughly 7-km increase in daily VKT per capita ($\beta = 7$, $p<0.001$), accounting for $22\%$ of explained variance.
$\Delta R_g$ ($\beta = 1.3$, $p < 0.01$) is the largest single contributor of explained variance at about $26\%$,
where a 1-km expansion in daily activity radius is associated with 1-km VKT increase.

Within urban form, baseline $PopGini$ ($\beta = -45$, $p < 0.001$) contributes about $22\%$ of explained variance.
For New York ($PopGini$ = 0.545) and Detroit ($PopGini$ = 0.266), compactness is associated with around 13-km less daily VKT per capita increase.
$UCI$ ($\beta = -29$, $p < 0.05$) contributes an additional $4\%$.
For Cleveland ($UCI$ = 0.175) and Sacramento ($UCI$ = 0.277), job concentration is associated with about 3-km less $\Delta$VKT.

To better understand how mobility behavior shifts factor into VKT growth, we also perform an arithmetic decomposition using the additive Laspeyres approach \cite{ang2004decomp, millardball2011dctravel}.
Since $VKT = T \cdot D$ (daily car trips per capita $\times$ average car trip distance), $\Delta VKT$ decomposes into a frequency effect ($\Delta T \cdot D_0$), a distance effect ($T_0 \cdot \Delta D$), and a small interaction term.
Comparing Fall 2024 VKT values to the Fall 2019 baselines across all 15 metros, we observe a positive frequency effect (adding about 8.8 km per capita per day), partially offset by a negative distance effect ($-$2.8 km due to shorter trips), yielding a net 5.2-km daily VKT increase per capita.

\section{Discussion}
\label{sec:discussion}
Post-pandemic mobility is shaped by hybrid work and space-time flexibility,
but characterizing this reorganization requires an approach that integrates multiple dimensions of urban travel.
This study develops a behavior-centric computational framework that examines post-pandemic mobility through sector composition, trip-level metrics, and mobility-derived spatial structure measures.
In particular, coupling trip centrality with work-trip attractor concentration disentangles commute and non-commute sources of travel demand.
This synthesis of compactness, mobility centrality, and commute anchoring moves beyond static proxies and provides a legible, replicable structural taxonomy for comparative urban analysis.
It contributes to longstanding debates in transportation planning and urban economics on the relationship between built environment and travel behavior.

Applied to population-scale data across 15 U.S. metropolitan areas spanning 2019--2024,
our approach shows elevated remote work with limited explanatory power from sector composition alone,
and increased daily car travel despite reduced commuting.
Trip frequency is the primary contributor to VKT growth, accompanied by contracted activity spaces and incomplete transit recovery.
These patterns appear across cities spanning a wide range of spatial structures.
The framework's urban form dimension makes this ubiquity interpretable:
coupling trip centrality with work-trip attractor reveals how the same behavioral shifts manifest through distinct configurations of commute and non-commute travel demand.

Cross-validation with independent data sources would strengthen these findings but faces two challenges.
Available benchmarks measure different quantities,
from recall-based travel surveys that underreport short trips \cite{wolf2003impact, bricka2012} to highway monitoring that aggregates all on-road vehicles without separating personal travel.
Available data also lack longitudinal consistency:
the most recent national travel survey changed sample design and collection methods, producing results not comparable with prior waves \cite{nhts2022compat};
census-derived commute data is lagged; and consistent time series from passive mobility sources remain difficult to assemble.
Cross-validation across cities and data sources remains necessary.
Open mobility data as a shared resource would serve both research and evidence-based planning.

\section{Methods}
\label{sec:methods}
\paragraph{Study area}
We select 15 metropolitan areas (See Supplementary Fig. S5) by the following criteria: (1) the most populous \cite{CityPopulationUSACombMetro}; (2) where mobility data is available; (3) represents diverse geographical coverage and urban contexts. Supplementary Tables S4-5 show their population and mobility data size. A Combined Statistical Area (CSA) merges two or more neighboring Core Based Statistical Areas (CBSA) with economic linkage and labor interdependence, defined by an employment interchange of $15\%$ or more measured through cross-commuting \cite{CRS_IF12704, DataGov_CBSA_2023}. We refer in shorthand the select CSAs by their largest composing cities (e.g., ``San Jose-San Francisco-Oakland'' by ``San Jose'').

\paragraph{Data}
\label{subsec:data}
\begin{enumerate}
    \item \textbf{Replica} Replica Places is an activity-based travel demand model that produces a complete, disaggregate trip table and synthetic population for a given metropolitan region, season, and day type (typical weekday or weekend) \cite{ReplicaSeasonalMobility,pozdnoukhov2018model}. The model operates at megaregion scale (10--50 million residents) with second-by-second temporal resolution and point-of-interest-level spatial resolution. The origin-destination (OD) of trips and home-work locations of users are provided as longitude-latitude at the census block group (CBG) centroid level.

    The model proceeds in four stages: 1)~Construct synthetic population from census microdata, 2)~Learn travel behavior models from mobile location data, 3)~Agent-based simulation of all trips on the transportation network, and 4)~Calibration against held-back ground truth including traffic counts, transit ridership, and transportation network company volumes. Outputs consist of three joinable tables (i.e., people, trips, and routes) of every synthetic person in the region (see Table~\ref{tab:input-data} for a summary of input data sources and methods).

    Subin et al. \cite{subin2024} validated tract and aggregate-level VKT derived from Replica against the Local Area Transportation Characteristics for Households Data (LATCH) national estimates \cite{Subin2024SupplementMethods, Subin2024SupplementData}. We also performed validations for the 15 CSAs in this study. Trip statistics (origin-destination flows, trip distance, mode, and departure time) align with LEHD Origin-Destination Employment Statistics LODES and  National Household Travel Survey patterns. CSA-level race, age and employment statistics align with the American Community Survey. Supplementary ``Replica Data Validation'' section provides full details. Replica data have also been applied in non-peer-reviewed transportation and land use studies \cite {brookings2023proximity,tomer2020ntb} and by multiple metropolitan planning organizations \cite{ReplicaValidation, ou2023ghg}.

    \item \textbf{Mobility}
    Replica provides trips for one weekday (Thursday) in Spring (representative of March, April, and May) and Fall (September, October, and November) across eight periods 2019--2024. It uses a consistent approach for seasonal mobility models to ensure comparability. Data is available for one weekday per season, two seasons per year, except 2019 and 2022 (only fall data available), and 2020 (unavailable). Travel behavior is derived from about 30 million mobile devices per month. Devices must have minimum 14-hour daily coverage, at least seven seasonal overnight stays, and at least 20 similar observed activity sequences; rare sequences are excluded for statistical robustness and privacy.

    Processed GPS traces are assembled into behavioral personas weighted to census population counts via iterative proportional fitting. Three models are applied sequentially: 1) an activity sequence model (ASM) generates the day's activity chain, 2) a location choice model (LCM) selects destinations from nearby venues ranked by popularity with Laplacian noise, 3) and a mode choice model (MCM) assigns travel mode via a utility function over travel time, cost, and traveler attributes.

    An agent-based mobility simulation executes resulting trips on a validated OSM road network with dynamic traffic assignment following Highway Capacity Manual speed--flow relationships, GTFS-based transit routing, and representations of non-motorized, freight, and pass-through traffic. Simulation outputs are calibrated against held-back ground truth (e.g., traffic counts averaged by day of week, transit ridership by line, and TNC volumes by zone) by adjusting route, mode, and transit choice parameters.

    \item \textbf{Demographics}
    Replica generates a synthetic population of households and persons by training two Bayesian networks on Census PUMS microdata, stratified by geographic area, household size, and person role. A convex optimization procedure reweights the generated profiles to match census aggregate totals at the block-group level across demographic attributes (e.g., household size, income, employment status, commute mode). 
    Employed persons are assigned work locations via CTPP home--work commute matrices supplemented by LEHD employment data and seasonal LBS adjustments; and school-aged residents are matched to schools by enrollment, district, and proximity. Vehicles are assigned per household using a K-nearest neighbors model trained on consumer marketing data.

    \item \textbf{Work-from-home} Replica models work-from-home status by combining monthly surveys conducted by Bureau of Labor Statistics (BLS) with mobile location data that captures time spent at home versus work, using the latter to update the BLS signal and distribute it spatially across Census Public Use Microdata Areas (PUMA).

    \item \textbf{Population density}
    The population with a spatial resolution of 30 arcseconds ($\sim 1$ sqkm resolution) is aquired from LandScan \cite{LandScan2025}.
\end{enumerate}

\newcolumntype{S}[1]{>{\linespread{1}\scriptsize\sffamily\raggedright\arraybackslash}p{#1}}

\begin{table}
\caption{Input data sources, processing methods, and outputs by core data product (CDP).}
\label{tab:input-data}
\centering
\setlength{\tabcolsep}{4pt}
\renewcommand{\arraystretch}{0.8}
\begin{tabular}{@{} S{1.6cm} S{2.0cm} S{3.2cm} S{4.8cm} S{3.0cm} @{}}
\toprule
\textbf{CDP} & \textbf{Source} & \textbf{Dataset} & \textbf{Method} & \textbf{Output} \\
\midrule
Population
  & Census Bureau & ACS/PUMS microdata & Bayesian nets (HH + person); convex optimization, 12 attributes & Synthetic HH and persons \\
  & Census Bureau & LEHD employment by sector & Sector-constrained assignment & Work locations \\
  & CTPP & Home--work flows (mode, income) & Matrix allocation; seasonal LBS adjustment & Commute O--D pairs \\
  & NCES/DOE & School enrollment & Proximity/enrollment-weighted assignment & School placements \\
  & Marketing vendor & Vehicle ownership (zip+2) & KNN (age, HH income) & Vehicle count/type per HH \\
  & State DMVs & BEV registration data & Weighted sampling; state-normalized & BEV share per HH \\
\midrule
\multirow{4}{*}{Trips}
  & \multirow{4}{*}{LBS aggregators} & GPS traces (30M+ devices/mo) & Space-time clustering; activity typing; IPF weighting & Staypoints, personas \\
  & & Activity day-sequences & Generative sequence model (ASM) & Activity chains \\
  & & Destination visit histories & Contextual ML model (LCM) & Destination choice \\
  & & Trip-level mode attributes & Discriminative ML + utility model (MCM) & Mode choice \\
\midrule
Built Environment
  & OSM Foundation & OSM road network & Verification (lanes, speeds, turns) & Routing network \\
  & Transit agencies & GTFS (300+ agencies) & Schedule parsing; 2$\times$/yr update & Transit timetables \\
  & \multirow{3}{*}{Multiple vendors} & Land use/parcel data & 10 L1, 18 L2 categories & Land use classes \\
  & & Building footprints & Composite modeling; ACS-scaled & Building area, dwellings \\
  & & POI directory (NAICS) & Monthly venue-level update & Venue locations \\
\midrule
Calibration
  & Sensors, gantries & Hourly traffic counts & Day-of-week averaging; outlier removal & Route/mode constants \\
  & Turnstiles, taps & Transit ridership per line & Algorithmic + manual QC & Transit parameters \\
  & TNC/taxi records & Zone-to-zone volumes & Verification filters & TNC parameters \\
\bottomrule
\end{tabular}

\smallskip\noindent\scriptsize\raggedright
\textit{Abbreviations:}
ACS, American Community Survey;
ASM, activity sequence model;
BEV, battery electric vehicle;
CTPP, Census Transportation Planning Products;
DOE, Department of Education;
GTFS, General Transit Feed Specification;
HH, household;
IPF, iterative proportional fitting;
KNN, K-nearest neighbors;
LBS, location-based services;
LCM, location choice model;
LEHD, Longitudinal Employer--Household Dynamics;
MCM, mode choice model;
ML, machine learning;
NAICS, North American Industry Classification System;
NCES, National Center for Education Statistics;
O--D, origin--destination;
OSM, OpenStreetMap;
POI, point of interest;
PUMS, Public Use Microdata Sample;
QC, quality control;
TNC, transportation network company.
\end{table}

\paragraph{Measuring job diversity}
Employment diversity is measured by applying the Shannon's Entropy Index across 20 NAICS job sectors in each CSA:
    \begin{equation}
    H(X) = -\sum_{i=1}^{n} p_i \log_2 p_i
    \end{equation}
where $p_i$ is the proportion of jobs in the $i$-th NAICS sector. Higher values indicate greater job diversity, with the maximum being 4.32, indicating perfectly even employment distribution.

\paragraph{Measuring mobility dynamics}
For each CSA and season, we compute aggregates of the following mobility measures from individual-level data. Since data is available for one weekday per season, the aggregates reflect daily behavior.

\begin{enumerate}
    \item \textbf{Vehicle kilometers traveled (VKT) per capita} measures the average distance traveled by car per person in the study area:
    \[
    \mathrm{VKT} = \frac{\sum_{i=1}^{N_v} d_i^{\text{car}}}{P},
    \]
    where \(d_i^{\text{car}}\) is the distance of car trip \(i\), \(N_c\) is the number of car trips, and \(P\) is the population.

    \item \textbf{Transit Kilometers Traveled (TKT) per capita}, similarly, is calculated by summing the distances of all transit trips and dividing by the population:
    \[
    \mathrm{TKT} = \frac{\sum_{i=1}^{N_t} d_i^{\text{transit}}}{P},
    \]
    where \(d_i^{\text{transit}}\) is the distance of transit trip \(i\), and \(N_t\) is the number of transit trips.

    \item \textbf{Car trips per capita} captures the daily car trip counts per person in the study area:
    \[
    \mathrm{CarTrips} = \frac{N_v}{P},
    \]
    where \(N_v\) is the total number of vehicle trips and \(P\) is the regional population.

    \item \textbf{Total trips per capita} captures the daily trip counts per person across all modes in the study area:
    \[
    \mathrm{Trip} = \frac{N_{\text{all}}}{P},
    \]
    where \(N_{\text{all}}\) is the total number of trips across all modes and \(P\) is the regional population.

    \item \textbf{Radius of gyration}, or $R_g$, is an established metric in human mobility literature \cite{gonzalez2008understanding, xu2023urbdyn}. It measures the spatial extent of an individual's mobility around their home location. For a sequence of \(n\) visited locations with coordinates \(l_i\) and home location \(l_h\):
    \[
    R_g = \sqrt{\frac{1}{n}\sum_{i=1}^{n}\left(l_i - l_h\right)^2},
    \]
    We average individual $R_g$ for the regional population to acquire the aggregate statistic.

\end{enumerate}

\paragraph{Measuring urban form}
We apply three existing measures of urban form, with the population-mobility framework proposed by Xu et al. \cite{xu2023urbdyn} and urban centrality by Pereira et al. \cite{pereira2013uci}.

\begin{enumerate}
    \item \textbf{Population Gini Index} ($PopGini$) captures how unevenly population density is distributed across 1-km grid cells with:
    \begin{equation}
        G = \frac{2}{n^{2}\bar{x}} \sum_{i=1}^{n} i (x_i - \bar{x}),
    \end{equation}
    where $x_i$ denotes population density in cell $i$, $\bar{x}$ is the citywide mean, and $n$ is the total number of cells. Higher values indicate a stronger concentration of population. Supplementary Fig. S11 provides full results.

    \item \textbf{Mobility Centrality Index}, or $\Delta$KS, quantifies mobility differentiation within 50-km from the CBD. For each 3-km concentric ring $r_i$, we compute the Kolmogorov-Smirnov (KS) distance between individual $R_g$ distribution in that ring and the distribution of CBD ring $r_0$:
    \begin{equation}
    KS(r_i \mid r_0) = \sup_{R_g} \left| F_{\langle R_g(r_i)\rangle}(R_g) - F_{\langle R_g(r_0)\rangle}(R_g) \right|,
    \end{equation}
    where $F_{\langle R_g(r)\rangle}$ denotes the empirical cumulative distribution function (CDF) in ring $r$. The slope of KS, or $\Delta$KS, thus shows the monocentricity gradient (Supplementary Fig. S9):
    \begin{equation}
    \Delta KS = \frac{d\, KS(r_i \mid r_0)}{d\, \tilde{r}},
    \end{equation}
    larger $\Delta$KS indicating greater mobility differentiation away from downtown. Supplementary Fig. S12 provides full results.

    \item \textbf{Urban Centrality Index} is measured using home-to-work trips, defined as work-purpose trips whose previous trip purpose is home, and computed using the \texttt{uci} package in R. Concentration of job attraction is quantified with the location coefficient
    \begin{equation}
    LC = \frac{1}{2} \sum_{i=1}^{n} \left| s_i - \frac{1}{n} \right|,
    \qquad s_i = \frac{E_i}{E},
    \end{equation}
    where $s_i$ is the share of all home-to-work trips ending in census block group (CBG) $i$, $E_i$ is the count of such trips attracted by CBG $i$, and $E$ is the total across all CBGs completely within 50-km radius from the CBD. Spatial proximity of attraction clusters is captured by:
    \begin{equation}
    V = S' D S, \qquad PI = 1 - \frac{V}{V_{\max}},
    \end{equation}
    where $S$ is the column vector of $s_i$, $D$ is the CBG centroid-to-centroid distance matrix, and $V_{\max}$ is the maximum attainable value of $V$ given the 50-km buffered area. The overall centrality measure ($UCI$) is thus
    \begin{equation}
    UCI = LC \times PI,
    \end{equation}
    which increases with both stronger concentration and tighter spatial clustering of employment destinations (Supplementary Fig. S10). Supplementary Fig. S13 provides full results.
\end{enumerate}

\paragraph{Decomposing VKT change}

We perform an arithmetic decomposition of VKT change using the additive Laspeyres approach \cite{ang2004decomp, millardball2011dctravel}. Daily per-capita VKT is the product of daily car trips per capita ($T$) and average car trip distance ($D = VKT / T$), such that VKT for a CSA ($c$) in a season ($s$) is:
\begin{equation}
    VKT_{cs} = T_{cs} \cdot D_{cs}.
\end{equation}
The change in VKT relative to the Fall 2019 baseline thus decomposes into three terms:
\begin{equation}
    \Delta VKT_{cs} = \underbrace{\Delta T \cdot D_0}_{\text{frequency effect}} + \underbrace{T_0 \cdot \Delta D}_{\text{distance effect}} + \underbrace{\Delta T \cdot \Delta D}_{\text{interaction}},
\end{equation}
where $T_0$ and $D_0$ are Fall 2019 baseline values, $\Delta VKT_{cs}$ denote changes in daily per-capita VKT for each CSA and season compared with Fall 2019 baselines. The three terms sum exactly to $\Delta VKT$ with no residual and no estimation. We compute this for each CSA and season, and report cross-city averages of Fall 2024 vs. Fall 2019 endpoints. Season-by-season results are provided in Supplementary Table S10.

\paragraph{Estimating VKT change associations}

We estimate an ordinary least squares model relating daily per-capita VKT changes to travel behavior changes and baseline urban-form characteristics:
\begin{equation}
    \begin{aligned}
    \Delta VKT_{cs}
    =\, & \beta_0 + \underbrace{\big( \beta_1 \Delta TKT_{cs}
                      + \beta_2 \Delta Trip_{cs}
                      + \beta_3 \Delta Rg_{cs} \big)}_{\text{mobility behavior change}} \\[4pt]
    & + \underbrace{\big( \beta_4 \Delta KS_{c19}
                      + \beta_5 PopGini_{c19}
                      + \beta_6 UCI_{c19} \big)}_{\text{baseline urban form}} \\[4pt]
    & + \underbrace{\beta_7 VKT_{c19}}_{\text{baseline travel demand}}
    + \varepsilon_{cs}.
    \end{aligned}
\end{equation}

$\Delta VKT_{cs}$ and $\Delta TKT_{cs}$ denote changes in daily per-capita car and transit distances traveled for each CSA and season compared with Fall 2019 baselines, $\Delta Trip_{cs}$ denotes changes in daily per-capita trips, and $\Delta Rg_{cs}$ denotes changes in average individual daily activity radius. $\Delta KS_{c19}$, $PopGini_{c19}$, and $UCI_{c19}$ denote CSA's baseline urban forms in Fall 2019, and $VKT_{c19}$ baseline daily per-capita VKT reflecting travel demand. The model's explained variance is decomposed using the Lindeman–Merenda–Gold (LMG) method \cite{Lideman1980}. For each variable $x_j$, the LMG value equals the mean increase in $R^{2}$ when adding $x_j$ to every subset of preceding predictors. Variable-level contributions are grouped by mobility behavior change ($\Delta TKT_{cs}$, $\Delta Trip_{cs}$, $\Delta Rg_{cs}$), baseline urban form ($\Delta KS_{c19}$, $PopGini_{c19}$, $UCI_{c19}$), and baseline travel demand ($VKT_{c19}$).

\newpage


\begin{figure}[htbp]
    \centering
    \includegraphics[width=\textwidth]{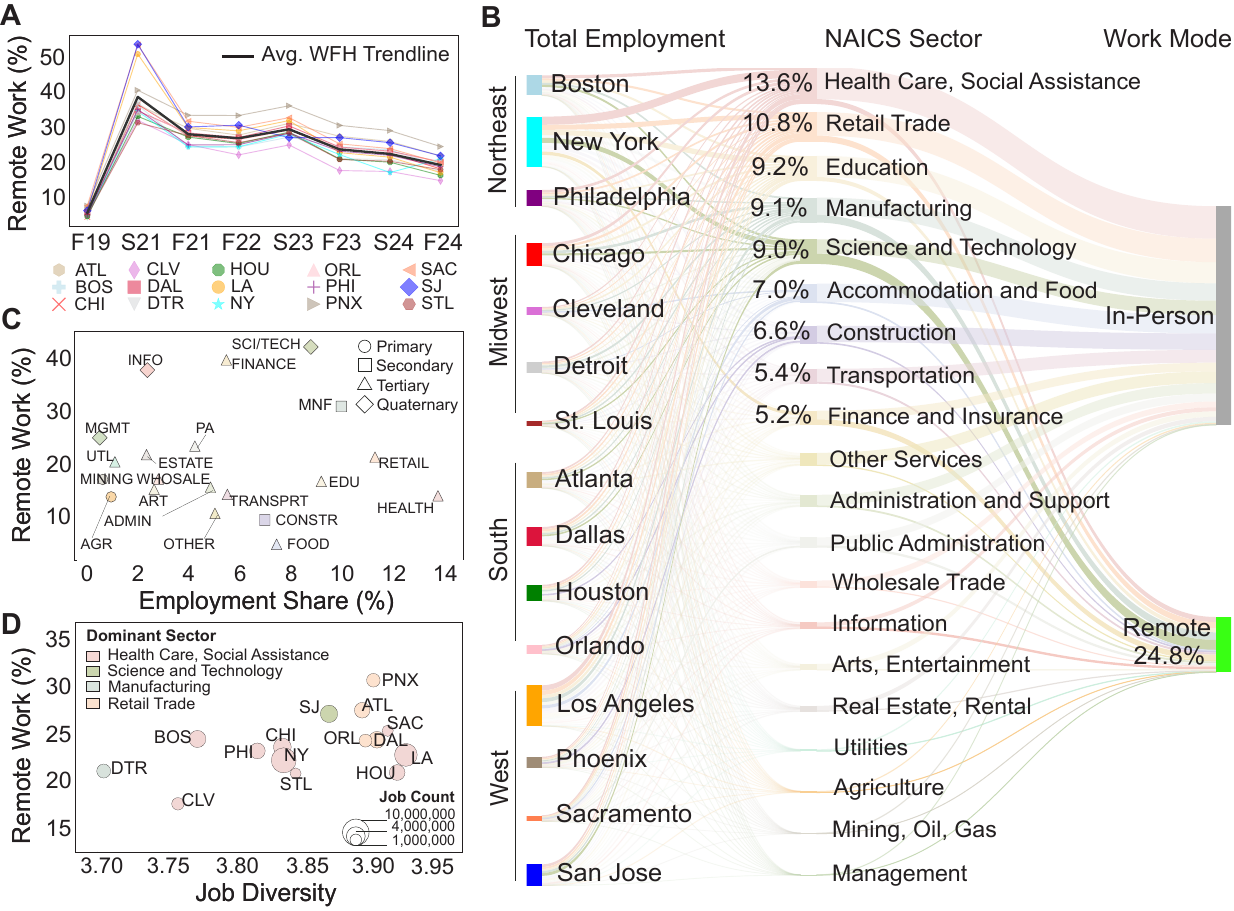}
    \caption{{\bf Status of remote work in 15 CSAs.} \textbf{A.} shows WFH trends from Fall 2019 to Fall 2024. Data was not available for 2020.
    \textbf{B-D.} use Fall 2023 data (See Supplementary Table S1 for details).
    \textbf{B.} traces all workers in the select CSA by labor market size (left panel), sectors (middle), and work mode (right).
    \textbf{C.} shows the detailed remote work vs. employment share breakdown of each sector by four main economic divisions.
    \textbf{D.} shows CSA's remote work share vs. job diversity. Job diversity is calculated via Shannon entropy of employment across sectors. CSA bubbles are sized by total jobs and colored by the dominant sector with the most employment (See Supplementary Fig. S7 for top sectors by CSA).}
    \label{fig: wfh}
\end{figure}

\begin{figure}[htbp]
    \centering
    \includegraphics[width=\textwidth]{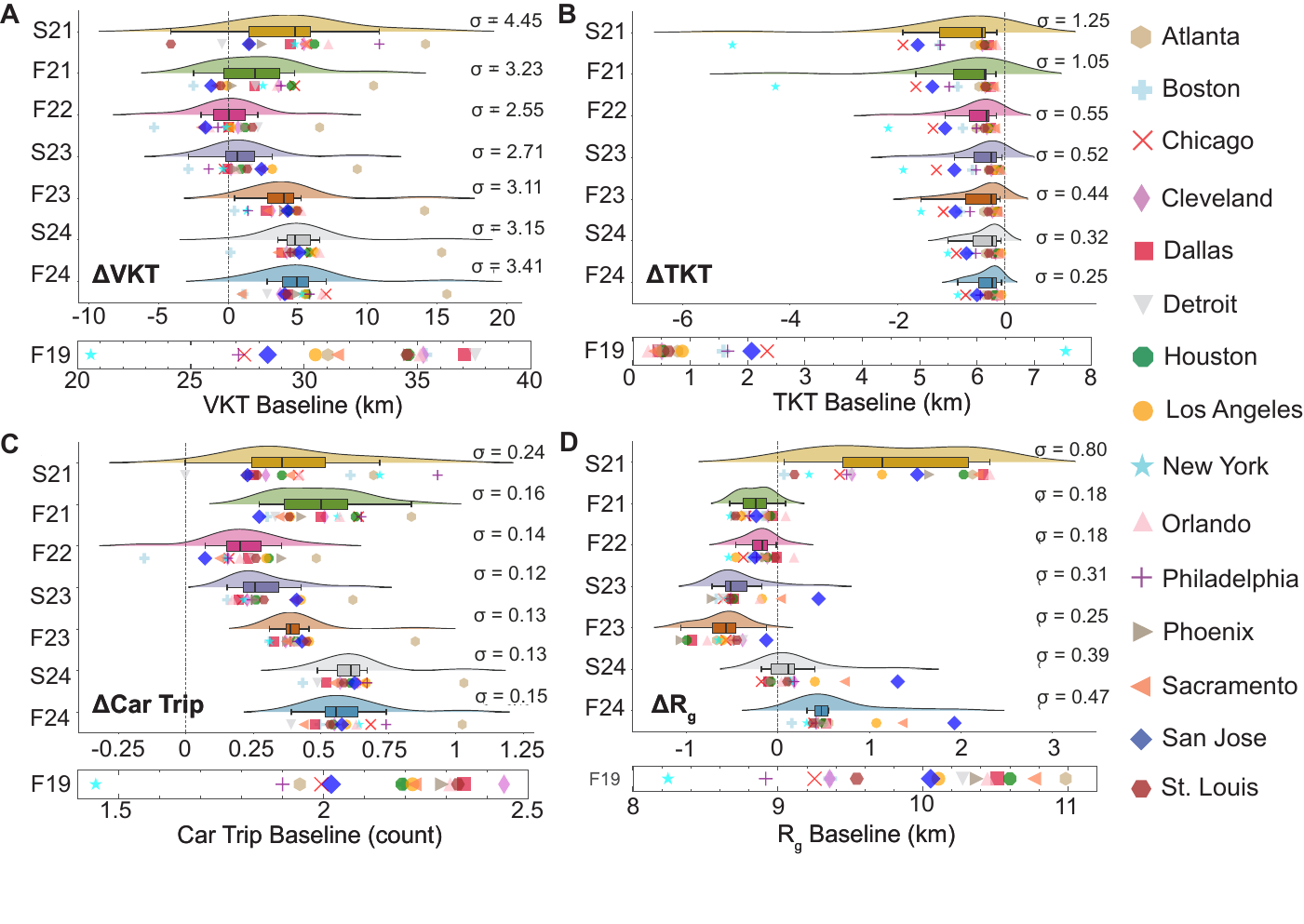}
    \caption{{\bf Mobility behavior in 15 CSAs from 2019 to 2024}. Changes are relative to Fall 2019 baseline values. Statistics are computed from individual-level data and aggregated at the CSA level (See Supplementary Table S5). \textbf{A.} Changes in daily vehicle kilometers traveled per capita ($\Delta$VKT). Baseline VKTs range 20--40 km. CSAs experienced around 5-km $\Delta$VKT post-pandemic. \textbf{B.} Changes in daily transit kilometers traveled per capita ($\Delta$TKT). Baseline TKTs range 0--8 km; denser CSAs have more transit usage and experienced more pronounced disruptions. \textbf{C.} Changes in daily car trips per capita ($\Delta$Car Trip). Baseline daily car trips per capita range from 1 to 3. On average, people make about 1 additional car trip every two days post-pandemic. \textbf{D.} Changes in daily individual radius of gyration ($\Delta R_g$), or activity radius, in km. Baseline $R_g$ values range 8--11 km. $R_g$ shrank in CSAs post-pandemic but showed signs of rebound in 2024.}
    \label{fig: raincloud}
\end{figure}

\begin{figure}[htbp]
    \centering
    \includegraphics[width=\textwidth]{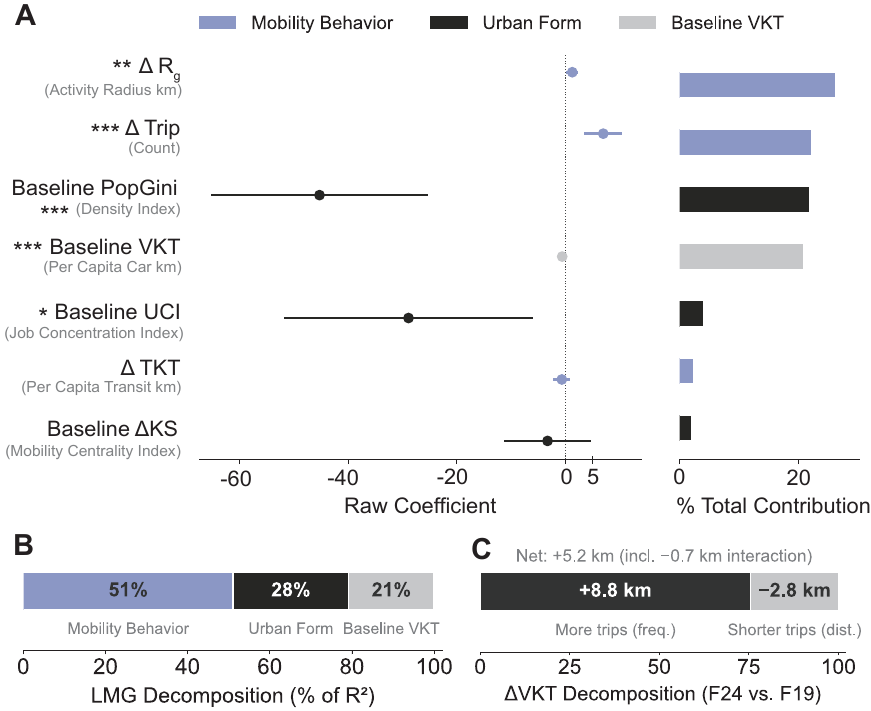}
    \caption{{\bf VKT growth is driven by increased trip frequency and moderated by compact urban form.} \textbf{A.} OLS coefficients (95\% CI) and individual LMG relative importance for $\Delta$VKT regressed on mobility behavior changes (vs.\ Fall 2019) and baseline urban form ($R^2 = 0.32$; N = 15 CSAs $\times$ 7 seasons). Trip frequency ($\Delta$Trip; $\beta = 7$, $p < 0.001$) and daily activity radius expansion ($\Delta R_g$; $\beta = 1.3$, $p < 0.01$) are the leading behavioral drivers. Compact urban form ($PopGini$; $\beta = -45$, $p < 0.001$) and employment concentration ($UCI$; $\beta = -29$, $p < 0.05$) significantly dampen VKT growth. \textbf{B.} Grouped LMG decomposition of $R^2$: mobility behavior $51\%$, urban form $28\%$, baseline VKT $21\%$. \textbf{C.} Arithmetic decomposition of mean $\Delta$VKT, Fall 2024 vs.\ Fall 2019 baseline ($N = 15$ CSAs). Increased trip frequency (+8.8 km) is partially offset by shorter trip distances ($-$2.8 km), yielding a net 5.2-km VKT increase per capita per day.}
    \label{fig: regression}
\end{figure}

\clearpage

\nolinenumbers

\paragraph{Data availability statement}
All data needed to evaluate the conclusions are described
in the paper. Replica data are used under institutional access provided by the University of California, Berkeley, and raw data cannot be made available. Processed data are available at GitHub (\href{https://github.com/humnetlab/NewNormal}{https://github.com/humnetlab/NewNormal}).

\paragraph{Code availability statement}
Codes to reproduce figures and analyses presented in this work are available at GitHub (\href{https://github.com/humnetlab/NewNormal}{https://github.com/humnetlab/NewNormal}).

\paragraph{Acknowledgements}
We thank Joe Castiglione and Tilly Chang from the San Francisco County Transportation Authority for valuable feedback and insight. This work was supported by the ITS-SB1 Berkeley Statewide Transportation Research Program.

\paragraph{Author contributions}
B.E.C and M.C.G. conceived the research. B.E.C and M.C.G. developed the methodology. B.E.C processed the data, implemented the methodology and performed visualization. B.E.C and M.C.G. analyzed the results. B.E.C prepared the original draft. B.E.C and M.C.G. edited and revised the manuscript. M.C.G. supervised the research.

\paragraph{Competing interests} The authors declare no competing interests.


\clearpage
\bibliographystyle{unsrtnat}
\bibliography{biblio}

\end{document}